\documentclass[11pt]{article}

\usepackage{url,cite}

\usepackage[latin1]{inputenc}
\usepackage[english]{babel}

\usepackage{amssymb,amsmath,amsthm,cancel,hyperref,graphicx,xcolor}
\usepackage{picinpar,graphicx,xypic}
\usepackage{booktabs}
\usepackage{mathrsfs}
\usepackage[font=small,labelfont=bf,margin=3mm,labelsep=period,tableposition=top]{caption}
\usepackage[a4paper, top=2.2cm, bottom=2.2cm, left=2cm, right=2cm, bindingoffset=0mm]{geometry}

\pdfoutput=1

\newcommand{\mh}{m_H}

\newcommand{\mz}{m_Z}

\newcommand{\sss}{\scriptscriptstyle\rm}

\newcommand{\Ord}{\mathcal{O}}

\newcommand{\as}{\alpha_s}
\newcommand{\muf}{\mu_{\sss F}}
\newcommand{\mur}{\mu_{\sss R}}

\let\originalleft\left
\let\originalright\right
\renewcommand{\left}{\mathopen{}\mathclose\bgroup\originalleft}
\renewcommand{\right}{\aftergroup\egroup\originalright}

\def\beq{\begin{equation}}  
\def\eeq{\end{equation}}
\def\({\left(}
\def\){\right)}
\def\[{\left[}
\def\]{\right]}

\begin{document}
\begin{flushright}
{\small
DESY 14-052\\
IFUM-1026-FT\\
DCPT/14/62\\
IPPP/14/31
}
\end{flushright}
\vspace{1.2cm}

\begin{center}
{\LARGE \bf Updated Higgs cross section at approximate N$^3$LO}
\vspace*{1.5cm}

{\bf Marco Bonvini}$^a$, {\bf Richard D.~Ball}$^{b}$, {\bf Stefano Forte}$^{c}$,\\ 
{\bf Simone Marzani}$^d$ and {\bf Giovanni Ridolfi}$^e$
\\
\vspace{0.4cm}
{\it
{}$^a$ Deutsches Elektronen-Synchroton, DESY,\\ 
Notkestra{\ss}e 85, D-22603 Hamburg, Germany\\ 
\medskip
{}$^b$Tait Institute, University of Edinburgh,\\  JCMB, KB, Mayfield Road, Edinburgh EH9 3JZ, Scotland\\ \medskip
{}$^c$Dipartimento di Fisica, Universit\`a di Milano and
INFN, Sezione di Milano,\\
Via Celoria 16, I-20133 Milano, Italy\\ 
\medskip
{}$^d$Institute for Particle Physics Phenomenology, Durham University,\\ South Road, Durham DH1 3LE, England\\ 
\medskip
{}$^e$Dipartimento di Fisica, Universit\`a di  Genova and INFN,
Sezione di Genova, \\Via Dodecaneso 33, I-16146 Genova, Italy
}
\vspace*{1.5cm}

\bigskip
\bigskip

{\bf \large Abstract:}
\end{center}

\begin{minipage}[c][][s]{0.9\textwidth}
We update our estimate of the cross section
for Higgs production in gluon fusion at
next-to-next-to-next-to-leading order (N$^3$LO) in $\as$ in view of 
the recent full computation of the result in the soft limit for
infinite top mass, which determines a previously unknown constant. We
briefly discuss the phenomenological implications.
Results are available through the updated version of the 
\texttt{\href{http://www.ge.infn.it/~bonvini/higgs/}{ggHiggs}} code.
\end{minipage}

\clearpage

The cross section for the production
of a Higgs boson in gluon-gluon fusion (the dominant production
subprocess at available collider energies) has been computed 
up to next-to-next-to-leading order (NNLO) in perturbative 
Quantum Chromodynamics (QCD)~\cite{higgsuptoNNLO},
and
a calculation of the N$^3$LO correction is under way~\cite{NNNLO,Anastasiou:2014vaa}.
In a recent paper~\cite{Ball:2013bra}
we have provided an estimate of the N$^3$LO correction,
based on the knowledge of the analytic structure of the coefficients
of the perturbative expansion
in the space of the variable $N$, Mellin conjugate to $z=\mh^2/\hat s$
where $\mh$ is the Higgs mass and
$\hat s$ is the squared partonic center-of-mass energy.
Such knowledge, in turn, originates from resummation of powers of
$\log N$ in the large-$N$ regime (soft-gluon or threshold resummation),
and from high-energy resummation, which fixes the behavior
of the coefficient in the vicinity of its rightmost singularity in the
$N$ complex plane. High-energy resummation turns out to have a relatively
small direct numerical impact, but its analyticity properties affect considerably 
the form of soft-emission logarithmic terms.

The result of Ref.~\cite{Ball:2013bra} in particular included all
contributions to the $\Ord(\as^3)$ which do not vanish as $N\to\infty$,
and which either grow logarithmically, or are constant. These contributions
were all known,
with the exception of the constant, which in $z$ space corresponds to
the coefficient of the $\delta(1-z)$ contribution to the
cross section. Recently~\cite{Anastasiou:2014vaa}, the full cross
section was determined in the 
soft limit, including this constant, and our approximate result can be
 updated accordingly: the result of Ref.~\cite{Anastasiou:2014vaa}
effectively amounts to a determination of the coefficient $g_{0,3}$ of
Ref.~\cite{Ball:2013bra}. The value of this coefficient 
was estimated in Ref.~\cite{Ball:2013bra} to be $g_{0,3}=114.7$ (for
finite $m_t$). This estimate was arrived at by rewriting $g_{0,n}=\bar
g_{0,n}+r_n$,  with $r_3$  known in terms of available
information, noting that the perturbative
behaviour of the known coefficients  suggests
$r_3\gg \bar g_{0,3}$, and thus simply assuming $g_{0,3}=r_3$.
Similar estimates for $g_0$ have also been obtained using methods for
the all-order resummation of constant contributions~\cite{PI_SQ}.

In Ref.~\cite{Anastasiou:2014vaa} the coefficient $g_{0,3}$ is
determined in the pointlike limit. We wish to use this result while
retaining the full $m_t$ dependence of all the remaining information
(which in particular is important in order to have the correct
analytic structure in the high-energy limit). Note that the $m_t$
dependence of this constant at previous orders is negligible: at NLO
the variation of $g_{0,1}$ when going from finite $m_t$ to the
pointlike limit is by about 0.5\%, and at NNLO the variation of 
 $g_{0,2}$ is by about 1\%. At N$^3$LO we have a certain latitude in
deciding which coefficient we should evaluate in the pointlike 
limit. We choose to take the pointlike limit value of the coefficient
of the delta function in the full $\Ord(\as^3)$ cross section (as given
in Eq.~(4) of Ref.~\cite{Anastasiou:2014vaa}). 

With this procedure we thus determine
$g_{0,3}=116.7$,
very close to the value   $g_{0,3}=114.7$ used  in
Ref.~\cite{Ball:2013bra}.
Using 
the resulting value for $\bar g_{0,3}=g_{0,3}-r_3=116.7-114.7=2.0$ 
in Eqs.~(4.1,4.2) of Ref.~\cite{Ball:2013bra}
we get the approximate updated
total cross section at the Large Hadron Collider (LHC), 
with $\sqrt s=8$~TeV and $m_H=125$~GeV,
computed using 
the NNLO NNPDF2.1~\cite{NNPDF21} set of parton distribution
functions
with $\as(\mz^2)=0.119$ (i.e.\ the same choices as in
Ref.~\cite{Ball:2013bra}) 
\begin{align}
\sigma_\text{approx}^{\text{N$^3$LO}}(s,\mh^2)
&= \big( 22.42 \pm 0.31  \big) \text{~pb} \qquad \text{for $\mur=\mh$}
\\
&= \big( 23.70 \pm 0.54  \big) \text{~pb} \qquad \text{for $\mur=\mh/2$},
\end{align}
where $\muf=\mh$, and
the error shown is our estimate of the uncertainty in our approximation.
The result is extremely
close to the value of Ref.~\cite{Ball:2013bra}, the difference being 
at the permille level.

\begin{figure}[t]
 \centering
 \includegraphics[width=0.95\textwidth,page=1]{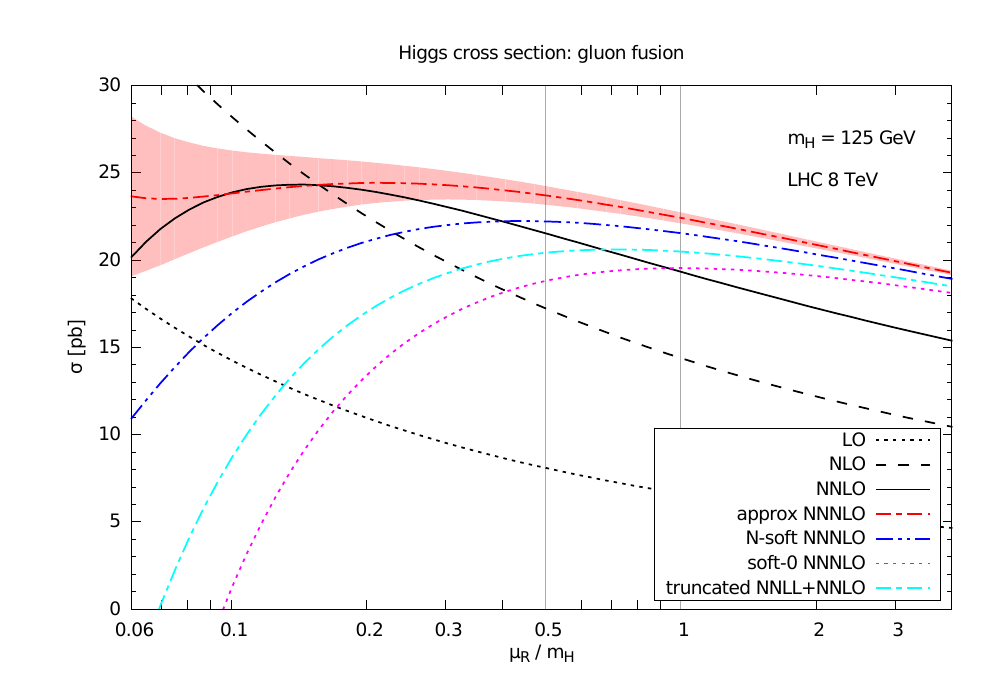}
 \caption{Dependence of the N$^3$LO cross section on
   the renormalization scale~$\mur$. Two common choices of renormalization scale
   are shown as vertical bars. The approximate N$^3$LO curves are,
   from top to bottom, our best approximation, the $N$-soft
   approximation, the N$^3$LO truncation of the NNLL resummed result of
   Ref.~\cite{deFlorian:2012yg}, and the soft-$0$ approximation (see text for details). 
   In all cases, the full result with finite top mass is included through NNLO.
   The known LO, NLO and NNLO results are also shown.
   The red band provides an estimate of the uncertainty on our result,
   obtained with  the procedure of Ref.~\cite{Ball:2013bra}.
 }
 \label{fig:hadronic_xs}
\end{figure}
Our result, and its dependence  on the renormalization scale $\mur$, are
shown in Fig.~\ref{fig:hadronic_xs} (the dependence on the
factorization scale is negligible, as discussed in
Ref.~\cite{Ball:2013bra}), compared to the lower-order results.
Our estimate of the N$^3$LO cross section is the red dot-dashed curve;
the band provides  our estimate for the uncertainty involved in the
approximation procedure,
details are given in Ref.~\cite{Ball:2013bra}.

We thus confirm, now on a firmer footing, the conclusion of
Ref.~\cite{Ball:2013bra}, namely that the N$^3$LO contribution leads
(for $\mur=\mh$) to an increase by about 16\% of the NNLO 
cross section.
Note that it was shown in Ref.~\cite{Forte:2013mda} that this conclusion
would very likely be unaffected by the consistent use of N$^3$LO parton
distributions.
This result can be compared to the main commonly-used
approximation to higher-order corrections, namely that from threshold
resummation at next-to-next-to-leading log (NNLL)~\cite{deFlorian:2012yg}. 
The truncation of the latter to
$\Ord(\as^3)$ is also shown in Fig.~\ref{fig:hadronic_xs}: it is
seen to lead to an increase of the NNLO by about 6\% at the same scale. Note that the (in
principle infinite) series of higher orders included in the
resummation only adds an extra 2\% to this.

This truncated NNLL
resummed result differs from
our approximation in three respects: the value of the constant (which
in Ref.~\cite{deFlorian:2012yg} corresponds to $g_{0,3}=0$); the
coefficient of the single-logarithmic term (both the constant and the
single log would only 
appear in next=to-next-to-next-to-leading log (N$^3$LL) resummation); and the fact that the constraints
due to matching to high-energy resummation and analyticity are not taken into
account. The effect of the
single logarithmic term is completely negligible, so the difference
is due in roughly equal proportion to each of the other two reasons. This
is also illustrated in Fig.~\ref{fig:hadronic_xs}: the $N$-soft
(see Ref.~\cite{Ball:2013bra} for the precise definition) curve
corresponds to using the exact constant (and single-logarithmic term),
but otherwise only including in the same form the N$^3$LO terms as in the
resummation (i.e. without matching and analyticity). This prediction is seen to
indeed lie half-way between our approximation and the truncated NNLL
resummed result.

Finally, we also show in Fig.~\ref{fig:hadronic_xs} the so-called
soft-$0$ approximation (again, see Ref.~\cite{Ball:2013bra} for a
precise definition). This basically amounts to only keeping soft
contributions, but in $z$ space rather than in $N$ space, and it would
predict a suppression, rather than an enhancement, of the N$^3$LO
cross section in comparison to the NNLO one, for a wide range of values of $\mur$. In the soft limit
this approximation coincides with the other approximations discussed
here, but away from the limit it differs from them by large
corrections suppressed by powers of $\frac{1}{N}$ [or $(1-z)$];  it
is known~\cite{Catani:2003zt,Ball:2013bra} to fail at NLO and NNLO,
essentially because it does not respect longitudinal momentum
conservation (albeit by subleading terms)~\cite{Catani:1996yz}. The
result found using this
soft-$0$ approximation was explicitly given  in
Ref.~\cite{Anastasiou:2014vaa}.

The updated prediction is available through the code
\texttt{ggHiggs} (version 1.9 onwards), publicly available at the website
\texttt{\href{http://www.ge.infn.it/~bonvini/higgs/}{http://www.ge.infn.it/$\sim$bonvini/higgs/}}.


\section*{Acknowledgments}
We acknowledge useful discussions with Claude Duhr and Franz Herzog on the results of Ref.~\cite{Anastasiou:2014vaa}.
The work of SM is supported by the UK's STFC. SF and GR are  
supported in part by an Italian PRIN2010 grant, and SF also by a
European Investment Bank  EIBURS grant, and by the European Commission through 
the HiggsTools Initial Training
Network PITN-GA-2012-316704.

\end{document}